\documentclass[11pt]{article}      
\pdfoutput=1
\usepackage{amsthm,amsmath,amssymb,xcolor,sectsty,latexsym,mathtools,hyperref,mathrsfs}
\usepackage{slashed}  
\usepackage{bm}
\usepackage{epsfig}
\usepackage{dcolumn}
\definecolor{vermelho}{cmyk}{0,.88,.77,.40}
\usepackage{cite}
\usepackage{feynmp}
\numberwithin{equation}{section}
\usepackage[textwidth=6.5in,top=1.2in,bottom=1.2in]{geometry}
\usepackage{setspace}
\chapterfont{\color{vermelho}}
\usepackage{simplewick}
\usepackage{hyperref}

\newcommand{\be}{\begin{equation}}
\newcommand{\ee}{\end{equation}}
\newcommand{\beq}{\begin{equation}}
\newcommand{\eeq}{\end{equation}}
\newcommand{\ba}{\begin{eqnarray}}
\newcommand{\ea}{\end{eqnarray}}
\newcommand{\bef}{\begin{figure}}
\newcommand{\eef}{\end{figure}}

\newcommand{\p}{\partial}
\newcommand{\al}{\alpha}

\newcommand{\g}{\gamma}

\newcommand{\cL}{{\cal L}}

\def\k{{\bf k}}

\newcommand{\nn}{\nonumber}

\newcommand{\f}{I}
\newcommand{\ff}{J}

\begin{document}

\thispagestyle{empty}
\begin{titlepage}
\nopagebreak

\title{  \begin{center}\bf General Relativity from Causality \end{center} }

\vfill
\author{Mark P.~Hertzberg$^{}$\footnote{mark.hertzberg@tufts.edu}, ~ McCullen Sandora$^{}$\footnote{mccullen.sandora@tufts.edu}}
\date{ }

\maketitle

\begin{center}
	\vspace{-0.7cm}
	{\it  $^{}$Institute of Cosmology, Department of Physics and Astronomy}\\
	{\it  Tufts University, Medford, MA 02155, USA}
	
\end{center}
\bigskip
\begin{abstract}
We study large families of theories of interacting spin 2 particles from the point of view of causality. Although it is often stated that there is a unique Lorentz invariant effective theory of massless spin 2, namely general relativity, other theories that utilize higher derivative interactions do in fact exist. These theories are distinct from general relativity, as they permit any number of species of spin 2 particles, are described by a much larger set of parameters, and are not constrained to satisfy the equivalence principle. We consider the leading spin 2 couplings to scalars, fermions, and vectors, and systematically study signal propagation in all these other families of theories. We find that most interactions directly lead to superluminal propagation of either a spin 2 particle or a matter particle, and interactions that are subluminal generate other interactions that are superluminal. Hence, such theories of interacting multiple spin 2 species have superluminality, and by extension, acausality. This is radically different to the special case of general relativity with a single species of minimally coupled spin 2, which  leads to subluminal propagation from sources satisfying the null energy condition. This pathology persists even if the spin 2 field is massive. We compare these findings to the analogous case of spin 1 theories, where higher derivative interactions can be causal. This makes the spin 2 case very special, and suggests that multiple species of spin 2 is forbidden, leading us to general relativity as essentially the unique internally consistent effective theory of spin 2.
\end{abstract}

\end{titlepage}

\setcounter{page}{2}

\tableofcontents

\newpage

\section{Introduction}

General relativity is beautifully consistent with all current observations over a fantastic range of scales. Although it is not UV complete, it does represent a consistent effective theory for energy scales well below the Planck scale. Here we would like to examine its theoretical underpinning. We are interested in whether there are universal principles upon which general relativity is built. Often one appeals to the equivalence principle, or the principle of minimal coupling, or the diffeomorphism symmetry or general co-ordinate invariance, or the idea of space-time curvature, or beauty, etc, to motivate the theory. However such principles do not have {\em universal} applicability. For instance, the equivalence principle evidently does not apply to other sectors of physics, such as electromagnetism or the strong force, which acts on different particles differently.  Minimal coupling is not universal in that particles such as neutrinos are not minimally coupled to photons, and the same is anticipated of hidden sector particles.

Nevertheless there are principles that appear universal, at least for energies below the Planck scale, namely Lorentz symmetry and quantum mechanics. It is known that these basic principles underpin the structure of the Standard Model of particle physics. There, the idea is simply to build a theory of a particular collection of particles of different spins and masses and focus on just the leading operators that survive at low energies, allowing some particles to be minimally coupled. In the case of gravitation, one might enquire as to whether only the principles of Lorentz symmetry and quantum mechanics are enough to uniquely specify the theory as being general relativity. Of course in this case the theory is non-renormalizable in 3+1 dimensions, but we only demand that we have a consistent effective theory. In fact it is often said that indeed general relativity is the unique theory of massless spin 2 particles at low energies. However, as was discussed in a recent paper \cite{Hertzberg:2016djj}  by one of us, and introduced in older work \cite{Wald:1986bj}, there are whole classes of other Lorentz invariant theories of massless spin 2 particles. Here one couples the linearized Riemann tensor directly to matter, as it automatically results in a gauge invariant interaction. In that recent paper \cite{Hertzberg:2016djj} the special case of coupling to photons was examined, where it was shown that there is superluminality of one of the polarizations of the photon.

In this paper we would like to perform a much more systematic analysis of all these other classes of spin 2 particles. We first demonstrate that by utilizing higher derivative interactions, one can build theories of an arbitrary number of species of spin 2, which couple with a large number of parameters. This makes the space of theories vastly greater than the special case of general relativity, which involves only a single minimally coupled spin 2 particle. We systematically lay out the various types of leading order interactions to matter, including coupling to scalars, fermions, and vectors. In each case we carefully examine whether there is some form of superluminality in either the matter particles or a spin 2 particle. Our basic findings are: (i) most interactions directly lead to superluminal propagation of either a spin 2 particle or a matter particle, (ii) interactions that are subluminal generate other interactions that are superluminal, and (iii) interactions that are subluminal and do not generate other interactions that are superluminal, do not represent true interacting spin 2 particles at all, but only self interactions in the matter sector. A summary of the theories considered and our findings is given in Table \ref{tableone}.

We contrast these results with the case of spin 1 particles, where the most leading order higher derivative couplings are perfectly causal. So the properties of spin 2 are quite distinct from spin 1. We argue that the size of superluminality in all these spin 2 theories can be made sufficiently large to lead to macroscopic time advance and problems with causality. We also point out that this cannot be readily fixed by the introduction of other operators. We contrast this to the case of general relativity, which leads only to subluminal propagation for matter in the presence of ordinary sources that satisfy the null energy condition. We also remark on similar problems that emerge in massive spin 2 theories. Finally we comment on self-interactions and operators of much higher dimension that can be causal, although they tend to be associated with the presence of ghosts.

Our paper is organized as follows: 
In Section \ref{space} we give the basic structure of theories of spin 2.
In Section \ref{Single} we study interactions involving a single graviton in couplings to matter.
In Section \ref{super} we show that these theories have superluminality.
In Section \ref{Multiple} we study interactions involving multiple gravitons in couplings to matter.
In Section \ref{General} we present a more general proof that these theories have superluminality.
In Section \ref{Comparison} we compare these findings of superluminality to general relativity, spin 1, and massive spin 2.
In Section \ref{Conclusions} we conclude.

\begin{table}[tb]
\vskip.4cm
\begin{center}
\begin{tabular}{|c|c|c|c|}
\hline Particles Involved & Interaction & Causal? & Real? \\
\hline
\multicolumn{4}{|l|}{One Spin 2}\\
\hline
Scalar & $R\phi^2$ & Yes &  No\\
 Scalar & $G_{\mu\nu}\partial^\mu\partial^\nu\phi $ & Depends on other sectors &  No\\
 Scalar & $ R^{\mu\nu\alpha\beta} \partial_\mu\p_\alpha\phi\partial_\nu\phi\partial_\beta\phi$ & No  & Yes\\
 Two Scalars  &  $R^{\mu\nu\alpha\beta}\p_\mu\phi\p_\nu\chi\p_\alpha\phi\p_\beta\chi$  &  No & Yes\\
Fermion   &  $R\bar\psi\bar\psi $ &  Yes &   No\\
Fermion &  $iR^{\nu\delta}\bar\psi\g_\nu\p_\delta\psi$  &  Depends on other sectors  &  No\\
Fermion & $R_{\mu\nu\alpha\beta}\bar\psi\sigma^{\mu\nu}\psi\bar\psi\sigma^{\alpha\beta}\psi  $ & Yes &  Yes (but generates $(\star)$)\\
Vector    & $R^{\mu\nu\alpha\beta} F_{\mu\nu}F_{\alpha\beta}  $     &  No   & Yes\\
\hline
\multicolumn{4}{|l|}{Two Spin 2}\\
\hline
Scalar &  $f(\phi)GB$ & No& Yes\\
Pseudoscalar &  $g(\phi)R\tilde R$ &  No & Yes\\
Fermion  & $\bar\psi\psi GB$ &  No    & Yes\\
Fermion $(\star)$  & $\bar\psi\gamma_5\psi R\tilde R$ & No  &Yes\\
\hline
\multicolumn{4}{|l|}{Two Massive Spin 2}\\
\hline
Scalar & $f(\phi)(h^2-h_{\mu\nu}h^{\mu\nu})$  &   No  & Yes\\
\hline
\multicolumn{4}{|l|}{Minimally Coupled Spin 2}\\
\hline
All &   General Relativity  &  Yes &  Yes\\
\hline
\multicolumn{4}{|l|}{One Spin 1}\\
\hline
Fermion &  $F_{\mu\nu}\bar\psi\sigma^{\mu\nu}\psi$ &Yes &Yes\\
Fermion & $iF_{\mu\nu}\bar\psi\gamma^\mu\partial^\nu\psi$ & Depends on other sectors & No\\
\hline
\multicolumn{4}{|l|}{Two Spin 1}\\
\hline
Scalar  & $f(\phi)F^{\mu\nu}F_{\mu\nu}$ & Yes & Yes\\
Pseudoscalar  & $g(\phi) F^{\mu\nu} \tilde F_{\mu\nu}$ & Yes & Yes\\
\hline
\end{tabular}
\end{center}
\caption{Table of the interactions considered in  paper. The columns indicate as follows: First is the particles involved in the interaction; Second is the explicit interaction term; Third is if the interaction leads to causality propagation, with the caveat that some depend on other sectors of the theory; Fourth is if the interaction produces a non-zero scattering amplitude and hence is ``real".}
\label{tableone}
\end{table}

\section{Space of Spin 2 Theories}\label{space}

In this work, we primarily focus on massless spin 2 particles; though we will discuss both spin 1 and massive spin 2 in Section \ref{Comparison}. Massless spin 2 particles must be comprised of 2 helicities in 3+1 dimensions to be associated with a unitary representation of the Lorentz group. We embed the particles into a tensor field $h_{\mu\nu,\f}$, where $\mu,\nu$ are Lorentz indices, and $\f$ is a species index, running from $\f=1,2,\ldots,N$ for $N$ species of spin 2 particles. This is a convenient way of describing local interactions, though it comes at the expense of introducing too many degrees of freedom. As is well known, this can only be fixed by the introduction of an identification into the theory. 

\subsection{Multiple Species}\label{MultipleSpecies}

At the linear level, we can introduce the following identification
\be
h_{\mu\nu,\f}\equiv h_{\mu\nu,\f}+\partial_\mu\alpha_{\nu,\f}+\partial_\nu\alpha_{\mu,\f}
\ee
where $\alpha_{\mu,\f}$ is a set of 4 arbitrary gauge functions for each species $\f$.
Upon gauge fixing, this means that $h_{\mu\nu,\f}$ is not a good Lorentz tensor, and so utilizing $h_{\mu\nu,\f}$ directly to build an interacting theory in a self-consistent manner is generically extremely difficult. We shall return to this shortly, but for now it is important to mention a possible way around this issue. By taking two derivatives of $h_{\mu\nu,\f}$ we can form the following manifestly gauge invariant and Lorentz covariant 4-tensor
\be
R_{\mu\nu\rho\sigma,\f}\equiv{1\over2}\left(\partial_\rho\partial_\nu h_{\mu\sigma,\f}+\partial_\sigma\partial_\mu h_{\nu\rho,\f}-\partial_\sigma\partial_\nu h_{\mu\rho,\f}-\partial_\rho\partial_\mu h_{\nu\sigma,\f} \right)
\label{LinearRiemann}\ee
which is proportional to the linearized Riemann tensor.

The Lorentz invariant theory of free spin 2 particles is of course a unique theory. It simply describes particles with helicity $\pm 2$ satisfying the dispersion relation $E=p$. At the level of a local Lagrangian it is therefore also unique (up to boundary terms and field re-definitions) and takes the form
\beq
\mathcal{L}_{\text{kin}}= \sum_{\f}\left[{1\over2}(\partial h_\f)^2-{1\over2}\partial h_{\mu\nu,\f}\partial h^{\mu\nu}_\f+\partial_\mu h^{\mu\nu}_\f\partial_\nu h_\f-\partial_\mu h^{\rho\sigma}_\f\partial_\rho h^\mu_{\sigma,\f}\right]\label{kin}
\eeq
Note that this is gauge invariant, up to boundary terms.

By exploiting the linearized Riemann tensor, interactions can be readily written down. If we consider an interaction involving only one spin 2 particle coupled to matter, it takes the form
\beq
\mathcal{L}_{\text{int}}=\sum_{\f}R_{\mu\nu\rho\sigma,\f}\tilde{T}^{\mu\nu\rho\sigma}_\f
\eeq
where $\tilde{T}^{\mu\nu\rho\sigma}_\f$ is any 4-index tensor built out of the matter degrees of freedom; ideally one that is constructed to be ghost free. Note that by integrating by parts, one can also express this as $h_{\mu\nu,I}$ coupled to an identically conserved tensor, but this formulation with Riemann makes the construction of gauge invariant operators more explicit. For example, this can be readily generalized to multiple gravitons in interactions by allowing for multiple insertions of the linearized Riemann tensor, which we shall return to in Section \ref{Multiple} (other work includes Refs.~\cite{Bai:2016hui,Bai:2017dwf}).

\begin{figure}[t!]
\center{\includegraphics[width=12cm]{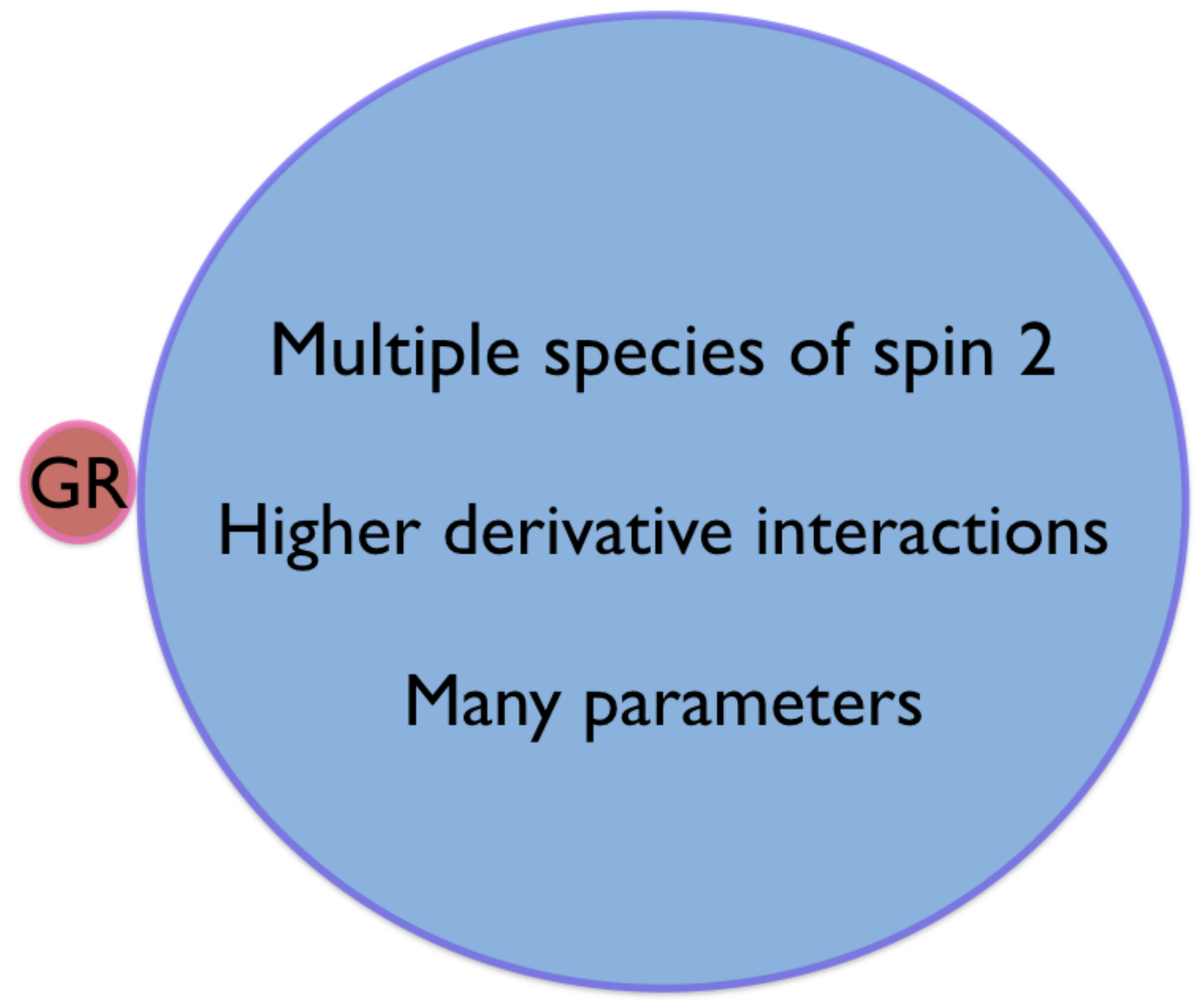}}
\caption{Space of possible theories of spin 2 particles. If only a single species of spin 2 is included with minimal coupling (requiring the nonlinear diffeomorphism invariance in its description), then we are uniquely led to general relativity (GR) at low energies; indicated by the small red circle on the left. If multiple species of spin 2 are included (requiring the linear gauge invariance in its description), which must be non-minimally coupled, then we are led to a vast array of possibilities associated with many parameters; indicated by the large blue circle on the right.  In this paper we show, however, that this much larger class of theories tend to suffer from problems of acausality, and thus are ultimately excluded from a physical viewpoint.} 
\label{SpaceOfTheories}\end{figure}

This allows for a huge set of theories, since there is tremendous freedom in the choice of $\tilde{T}_\f$; in fact this can be an arbitrary function, containing much more freedom than the usual coupling constant that specifies how a graviton couples to the matter Lagrangian. We shall systematically study couplings to scalars, fermions, and vectors in Section \ref{Single}. A concrete example is the operator
\be
\mathcal{L}_{\text{int}}={1\over\Lambda^3}\sum_{i\f} c_{i\f}\,R_{\mu\nu\rho\sigma,\f}F^{\mu\nu}_iF^{\rho\sigma}_i+\ldots
\ee
when coupling to vectors (the dots indicate other terms needed for consistency, that will be discussed in Section \ref{Vector}). Here the $i$ index runs over vector species and the $\f$ index runs over spin 2 species, so that there is a {\em matrix} of couplings $c_{i\f}$.
Hence this class of theories is associated with a huge range of parameters associated with an arbitrary number of species of spin 2. We have depicted this huge space of theories schematically as the blue region in Figure \ref{SpaceOfTheories}. Note that nothing in this framework demands the universality of couplings principle, so we allow the couplings to be arbitrary. This poses a severe challenge to derive the equivalence principle, which we aim to tackle in this work.

\subsection{Single Species}

The above analysis applies to any number $N$, including the special case of a single species $N=1$. However in this special case, a more relevant interaction is allowed by the Lorentz symmetry. All of the above interactions involve higher derivatives compared to so-called minimal coupling, where the $h_{\mu\nu,\f}$ is directly used. This is a well studied subject, which we recap only briefly here. 

To utilize minimal coupling, we would attempt to couple the spin 2 field directly to matter as follows
\be
\mathcal{L}_{\text{int}}=-{1\over2}\sum_\f \kappa_\f\,h_{\mu\nu,\f}T^{\mu\nu}_\f
\ee
This evidently violates the linear gauge invariance, unless $T^{\mu\nu}_\f$ is conserved. However, there is no non-trivially conserved tensor. The closest object is the matter energy-momentum tensor, which is only conserved in the limit in which we ignore this new interaction. 

Nevertheless we can proceed order by order in powers of the coupling $\kappa$. At second order in the coupling it is found that there is no consistent way to build the interactions without introducing an update to the gauge transformation rule. In fact, at all orders the gauge transformation is necessarily lifted to the full nonlinear diffeomorphism invariance
\be
\kappa\,h_{\mu\nu}(x^\beta)\equiv (\eta^\rho_\mu+\kappa\,\partial^\rho\alpha_\mu)(\eta^\sigma_\nu+\kappa\,\partial^\sigma\!\alpha_\nu)
(\eta_{\rho\sigma}+\kappa\,h_{\rho\sigma}(x^\beta+\kappa\,\alpha^\beta))-\eta_{\mu\nu}
\ee
Since there are only 4 coordinates, this only acts to remove 4 degrees of freedom, and hence this only works for a {\em single} species of spin 2. This conclusion is compatible with the traditional no-go theorems dictating the inconsistency of multiple interacting spin 2 fields \cite{Boulanger:2000rq,Benincasa:2007xk} where minimal, or leading order, couplings are assumed. In contrast, in the previous subsection this conclusion is avoided because the linearized Riemann tensor is used, which is explicitly invariant under the gauge identification. 

Furthermore, the entire action is determined to all orders uniquely, up to boundary terms and field redefinitions, in terms of the single coupling $G_N=\kappa^2/(16\pi)$, giving rise to the Einstein-Hilbert action \cite{Deser:1969wk}
\be
S=\int d^4 x\sqrt{-g}\left[{\mathcal{R}\over 16\pi G_N}+\mathcal{L}_M(\psi_i,g_{\mu\nu})\right]
\ee
where the script $\mathcal{R}$ is the full non-linear Ricci scalar (and we reserve the straight ``$R$" to refer to the canonically normalized linearized piece only as in eq.~(\ref{LinearRiemann})). This essentially unique theory is summarized by the small red circle in Figure \ref{SpaceOfTheories}. We do note that we can add to this action higher dimension operators of the form we summarized in the previous subsection, by exploiting the (nonlinear) Riemann tensor, and so other sub-leading interactions are allowed. However, since it necessarily only involves a single spin 2 species, it is much more restricted and associated with far fewer parameters than the other set of theories described above. 

Famously, this leads to the $1/r^2$ force law, the universality of free-fall, and all the successes of general relativity. At the same time this theory carries various conceptual puzzles, such as the notorious cosmological constant problem and the black hole information paradox. In this paper we would like to provide a deeper explanation as to why nature has nevertheless chosen this special theory, rather than the much bigger space of theories depicted in Figure \ref{SpaceOfTheories}.

\section{One Spin 2 in the Interaction}\label{Single}

Our focus in this and the next few sections is to develop and examine the large class of theories of Section \ref{MultipleSpecies} which may involve any number of spin 2 particles.

We first investigate interactions with matter that only involve one spin 2 particle in the interaction vertex. We consider explicit interactions with scalars, fermions, and vectors in turn.  The boson interactions will lead to novel, nontrivial theories that prima facie appear perfectly healthy.  However, in section \ref{super} we will show that all of these lead to superluminal propagation of matter, and ultimately acausality.  There appears to be at least one fermion interaction that evades this problem, but in Section \ref{fermRR} we will explain why it too is associated with a form of superluminality.

\begin{centering}
\begin{figure*}[h]
\centering
\includegraphics[width=15cm]{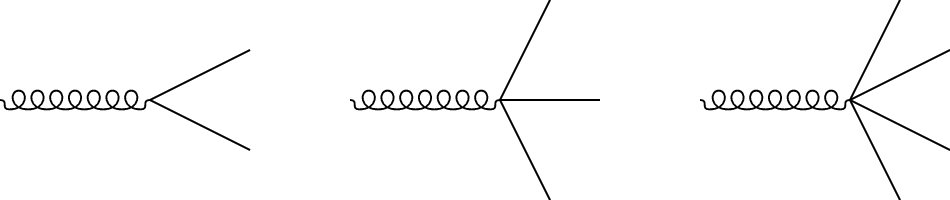}
\caption{The diagrams corresponding to the theories considered in this section.  The straight lines represent matter, and the squiggly lines represent spin 2 particles.}
\label{tree1}
\end{figure*}
\end{centering}

The Feynman diagrams for the interactions considered here are depicted in Fig.~\ref{tree1}.  Since in this section we shall only make reference to a {\em single} spin 2 particle in the interaction, we shall suppress the species index $\f$, though the extension to a sum over multiple species is straightforward.

\subsection{One Scalar}

The most general interaction Lagrangian linear in the spin 2 field and involving no more than one derivative on the scalar is
\beq
\cL_{\text{int}}=a_1Rf(\phi)+\frac{a_2}{\Lambda^3}R(\partial\phi)^2+\frac{a_3}{\Lambda^3}G^{\mu\nu}\p_\mu\phi\p_\nu\phi\label{scal}
\eeq
In general the coefficients $a_i$ can be functions of the scalar, but to lowest order we can take them to be constant.  
The first term will not affect the causality properties of the scalar, since it acts as a potential term and is unimportant in the large momentum limit.  The other two terms involve derivatives of the scalar, and so can potentially lead to alterations of the causal structure of the theory.  However, these interactions are fictitious\footnote{Also, note that the theory contains Ostrogradski ghosts unless $a_2=0$ \cite{Deffayet:2011gz}.}.  This can be seen by looking at the scattering amplitude for
\be
h+\phi\to h+\phi
\ee 
in the gauge $\partial^\mu h_{\mu\nu}=h^\mu_\mu=0$, in which case $G_{\mu\nu}=\Box h_{\mu\nu}$.  In this gauge, the interaction vertex is proportional to $p^2 \epsilon(p)_{ij}k^i_1k^j_2$, and vanishes when the external spin 2 field is placed on shell, where $p^2=0$.  Similarly, any scattering built out of this vertex vanishes when one of the spin 2 fields is on shell.  This is a direct consequence of the fact that this interaction is proportional to the free field equations of motion, which are used to define the Fock space in the interaction picture.  Consequently, this term can be completely removed by a field redefinition, and so is a `redundant operator'.  Notice that this removal is exact for our theories because the Einstein tensor is exactly linear in the spin 2 field.  This is in contrast to general relativity, where the Einstein tensor contains higher order terms, which lead to couplings that are not removed by the redefinition provided.

So although terms like $\sim R\,f(\phi)$ are perfectly causal, they do not actually involve the spin 2 particle at all. Other terms like $R(\partial\phi)^2$ are ambiguous in that their (a)causality depends on other sectors of the theory, but they also do not actually involve the spin 2 particle. This will be an important observation for all of the theories we discuss in this work, and so we show explicitly how a field redefinition transforms these interactions away in full generality in the Appendix.  

In order for the interaction to be nontrivial, then, it must involve the (linearized) Riemann tensor.  It is impossible to contract Riemann with a quantity involving first derivatives of a single scalar field, but if we consider terms involving second derivatives we can write 
\beq
\cL_{\text{int}}=\frac{a_4}{\Lambda^6}R^{\mu\nu\alpha\beta}\,\p_\mu\phi\,\p_\alpha\phi\,\p_\nu\p_\beta\phi+\ldots\label{R112}
\eeq
The term written explicitly has fourth (time) derivative equations of motion and so will contain Ostrogradski ghosts, but if we add terms proportional to the Einstein tensor to the Lagrangian with the right coefficients the equations can be made second order \cite{Deffayet:2011gz}.  Alternatively, we may use the second Bianchi identity to write the divergence of the Riemann tensor appearing in the equations of motion in terms of the Einstein tensor, $\p^\g R_{\al\beta\g\delta}=\p_\alpha R_{\beta\delta}-\p_\beta R_{\al\delta}$, which will vanish on flat backgrounds.  The equation of motion for the scalar field becomes
\beq
\Box\phi-\frac{a_4}{\Lambda^6}R^{\mu\nu\alpha\beta}\,\p_\mu\p_\al\phi\,\p_\nu\p_\beta\phi+\ldots=0\label{eom0}
\eeq
Here we have only shown the terms with the most derivatives acting on the scalar field, which amounts to treating the Riemann field as slowly varying compared to the scalar.  In section \ref{super} we show that this leads to superluminality on nontrivial backgrounds.

\subsection{Two Scalars}
We can also construct an interaction with the Riemann tensor using only first derivatives if we are willing to introduce a second scalar field.  The interaction is:
\beq
\cL_{\text{int}}=\frac{a_5}{\Lambda^7} R^{\mu\nu\al\beta}\,\p_\mu\phi\,\p_\nu\chi\,\p_\al\phi\,\p_\beta\chi
\eeq
As with the interaction (\ref{R112}) before, this term does not introduce ghosts in the scalar sector on flat backgrounds, and the ghosts can be removed completely if terms proportional to the Einstein tensor are added \cite{Ohashi:2015fma}.

The scalar equations of motion for this theory are (again, keeping only the leading derivative terms),
\begin{equation*}
\left(
\begin{array}{ccc}
\eta^{\mu\nu}-\,t_{\chi\chi}^{\mu\nu} & -\,t_{\chi\phi}^{\mu\nu}  \\
-\,t_{\chi\phi}^{\mu\nu} & \eta^{\mu\nu}-\,t_{\phi\phi}^{\mu\nu}
\end{array} \right) \p_\mu\p_\nu \left(
\begin{array}{ccc}
\phi \\ \chi
\end{array} \right)=0\label{eom00}
\end{equation*}
Here, we use the notation $t_{\psi_1\psi_2}^{\mu\nu}=2a_5/\Lambda^7R^{\mu\alpha\nu\beta}\p_\al\psi_1\p_\beta\psi_2$.  To show that this system exhibits superluminal motion, it suffices to only take one of these backgrounds to be nontrivial.  In this instance, the kinetic matrix diagonalizes and one of the fields is dictated by the wave equation.  This background is investigated in section \ref{super}.

\subsection{Fermions}

We now ask whether it is possible to couple fermions to the Riemann tensor, or if these types of couplings lead to superluminality as well.  At cubic order in the fields the most general form of the interactions can be written as
\ba
\cL_{\text{int}}&=&\frac{1}{\Lambda^2}\left(b_1R\bar\psi\psi+b_2R_{\al\beta\g\delta}\bar\psi\sigma^{\al\beta}\sigma^{\g\delta}\psi\right)\nn\\&+&\frac{1}{\Lambda^3}\left(ib_3R\bar\psi\g^\mu\p_\mu\psi+ib_4R^{\nu\delta}\bar\psi\g_\nu\p_\delta\psi+ib_5R_{\al\beta\g\delta}\bar\psi\sigma^{\al\beta}\g^\g\p^\delta\psi\right)\label{Rpsipsi}
\ea
Here we have neglected parity violating terms involving $\g_5$, but these do not alter our conclusions.  The first line involves no derivatives on the fermion, and so will not alter the causality structure of the theory.  However, both of these terms are trivial.  The first one is proportional to the Einstein tensor, so, as before, can be removed with a field redefinition.  The second potentially benign term involves the Riemann tensor and so is a genuine interaction, and resembles a mass term for the fermion on a given spin 2 background.  However, this term can be shown to vanish.  This is because the two sigma matrices can be replaced with their symmetrization, as a consequence of the index structure of the Riemann tensor.  Then the following gamma matrix identity can be employed:
\beq
\{\sigma_{\mu\nu},\sigma_{\alpha\beta}\}=2i(\eta_{\mu\alpha}\sigma_{\nu\beta}-\eta_{\mu\beta}\sigma_{\nu\alpha}-\epsilon_{\mu\nu\alpha\beta}\g^5)\label{anticomm}
\eeq
When contracted with the Riemann tensor, this becomes
\beq
R^{\mu\nu\alpha\beta}\{\sigma_{\mu\nu},\sigma_{\alpha\beta}\}=4iR^{\nu\beta}\sigma_{\nu\beta}-2i\epsilon_{\mu\nu\alpha\beta}R^{\mu\nu\alpha\beta}\g^5
\eeq
The first term vanishes because it is a symmetric tensor contracted with an antisymmetric tensor, and the last vanishes by the Bianchi identity.

The second line of (\ref{Rpsipsi}) involves couplings containing derivatives acting on the spinor field.  The first two terms are proportional to the Einstein tensor, and were shown in \cite{Ohkuwa:1980jx} to arise from integrating out heavy gauge bosons in the standard model.  There, it was shown that these can lead to superluminality in particular backgrounds, but these will not contribute in the absence of a source for the spin 2 field.  The last term can also be reduced to this form as well by employing spinor identities:
\ba
\Delta\cL_{\text{int}}&=&i\frac{b_5}{\Lambda}R_{\al\beta\g\delta}\bar\psi\sigma^{\al\beta}\g^\g\p^\delta\psi\nn\\
&=&\frac{b_5}{\Lambda}R^{\mu\nu\g\delta}\bar\psi(-\eta_{\mu\g}\g_\nu+\eta_{\nu\g}\g_\mu-\epsilon_{\mu\nu\g\sigma}\g^\sigma\g^5)\p_\delta\psi\nn\\
&=&\frac{2b_5}{\Lambda}R^{\nu\delta}\bar\psi\g_\nu\p_\delta\psi
\ea
Thus, even at the single derivative level, all cubic fermion couplings can be removed with a field redefinition.  If we are willing to consider higher powers of the fields, however, it is possible to write down nontrivial interacting terms.  The first of these is a dimension 9 operator:
\beq
\cL_{\text{int}}=\frac{b_5}{\Lambda^5}R_{\mu\nu\alpha\beta}\bar\psi\sigma^{\mu\nu}\psi\bar\psi\sigma^{\alpha\beta}\psi\label{Rpsi4}
\eeq
While this term is benign from the standpoint of causality, through loop effects it will generate dangerous terms that will be considered in section \ref{fermRR}.

\subsection{Vector}\label{Vector}

We now turn to interactions between a spin 1 and spin 2 field.  Of the possible cubic interactions, there most general (parity-even) Lagrangian is
\beq
\cL_{\text{int}}=\frac{1}{\Lambda^3}(c_1RF^2+c_2R^{\mu\nu}F_{\mu}{}^\lambda F_{\lambda\nu}+c_3R^{\mu\nu\al\beta}F_{\mu\nu}F_{\alpha\beta})
\eeq
This was first discussed in the context of the low energy effective theory for quantum electrodynamics in curved spacetime in \cite{Drummond:1979pp}.  (For a thorough discussion see \cite{Shore:2003zc}).  As before, the first two terms do not represent true interactions, and they vanish on backgrounds that are Ricci flat, so we focus on the last.  The equation of motion for the spin 1 field is
\beq
\p_\al F^{\al\beta}-4\frac{c_3}{\Lambda^3}R^{\mu\nu\alpha\beta}\p_\alpha F_{\mu\nu}\label{eom1}+\ldots=0
\eeq
(again we have only shown the terms that dominate when the background field is slowly varying).
The characteristics for the spin 1 field set by this last equation will be analyzed in the next section, and lead to superluminality.

\section{Signal Propagation}\label{super}

The equations of motion for the three interacting theories we have uncovered so far (\ref{eom00}), (\ref{eom0}), and (\ref{eom1}) all have the same basic form.  We will now show that all these admit superluminal propagation while remaining in the regime of validity of the effective field theory.  Using the eikonal (geometric optics) approximation to the equations of motion, where the wavelength of the matter particle is much smaller than the scale at which the background varies, we get that the characteristics obey
\be
(\eta^{\mu\nu}- R^{\mu\nu\alpha\beta}\chi_{\nu\beta})k_\mu k_\nu=0
\eeq

The only difference between the three theories is the form of the tensor $\chi^{\al\beta}$, which for the various theories we consider is  
\begin{equation}
\chi_{\nu\beta}= \left\{
 \begin{array}{rl}
  \frac{a_5}{\Lambda^6}\p_\nu\p_\beta\bar\phi & \text{scalar}\\
  \frac{2a_6}{\Lambda^7}\p_\nu\bar\chi\p_\beta\bar\chi & \text{two scalars}\\
   \frac{2c_3}{\Lambda^3}\epsilon_\nu\epsilon_\beta & \text{vector}
 \end{array} \right.
\end{equation}
In the first case the tensor is the second derivative of a background field, but if this is a plane wave it becomes $\chi_{\mu\nu}=-a_5p_\mu p_\nu\phi_p/\Lambda^6$.  The $p_\mu$ vectors are lightlike if the scalar is massless, and timelike if the scalar is massive.  The same holds for the second case, with $\phi_p\rightarrow\chi_p^2$.  In the last case, the polarization vectors are spacelike.  Generically, then, we make the replacement $\chi_{\nu\beta}\rightarrow qp_\nu p_\beta$.  

\subsection{Superluminality}

We now show that theories of this type exhibit superluminality on certain backgrounds.  Let us take the Riemann tensor to be a plane wave, so that it is a solution to the free field equations of motion $G_{\mu\nu}=0$.  Then
\beq
R^{\mu\alpha\nu\beta}\chi_{\alpha\beta} k_\mu k_\nu= q\left(h_{kp},{}_{kp}-\frac12h_{kk},{}_{pp}-\frac12h_{pp},{}_{kk}\right)\label{honk}
\eeq
Where subscripts denote projections, e.g. $h_{kp}=h_{\mu\nu}k^\mu p^\nu$, and subscripts after commas denote derivatives.

To be explicit, let's now specialize to a particular setup:  we will take $\gamma$ to propagate in the $\hat{z}$ direction, and will take $\vec{k}$ and $\vec{p}$ to be perpendicular to this direction, say $\vec{p}=|p| \hat{x}$.  In the case where $p^\mu$ is null and future directed, then $p^0=|p|$, and so our characteristic equation reduces to
\beq
-\omega^2+k^2+q|p|^2\left(\omega\ddot{h}_{x j}k_j-\frac12\ddot{h}_{ij}k_ik_j-\frac12\omega^2\ddot h_{xx}\right)=0
\eeq
This will define a dispersion relation $\omega(k)$, which will set the speed of propagation of the system through $\omega=vk$.  For simplicity we specialize further to keeping only the cross polarization of the spin two field, $h_\times=h_{xy}$, arriving at 
\begin{equation*}
\tilde{\omega}^2=\left(k_x\,\, k_y\right)\left(
\begin{array}{ccc}
1 & -\kappa/2  \\
-\kappa/2 & 1+\kappa^2/4 
\end{array} \right)  \left(
\begin{array}{ccc}
k_x \\k_y
\end{array} \right)
\end{equation*}
We have introduced the quantity $\kappa=q|p|^2\ddot\gamma_\times$ and shifted the frequency variable by $\tilde\omega=\omega-\kappa k_y/2$, which accounts for the fact that all signals drift uniformly in the $y$ direction.  From here the speed can be read off by looking at the eigenvalues of this matrix.  The trace and determinant condition give
\beq
v_1^2+v_2^2=2+\frac14\kappa^2,\quad v_1^2v_2^2=1
\eeq
One of these will necessarily be greater than 1 unless $\kappa=0$, since the determinant is unchanged.  

If the vector $p^\mu$ is spacelike (relevant to the spin 1 case) the analysis is even easier: since we took it to be perpendicular to the direction of propagation of the spin 2 field, the derivatives in equation \ref{honk} must be contracted with $k^\mu$.  Then we have
\beq
(1+q\chi'^2\ddot h_{+})\omega^2=k^2
\eeq
so  the speed can be either greater than or less than the speed of light, depending on the sign of $\ddot h_+$.

Now, to make the setup more explicit, we imagine we have a source for the spin 2 field, generating lots of fairly low energy particles.  The ultimate strength of this effect will depend on the intensity of the wave, which can be made arbitrarily high, and the wavelength, which can be made arbitrarily wide.  We wish to send a particle through this wave perpendicularly, all with the entire setup in a background $\chi$ field of very long wavelength.  With this setup the magnitude of the effect can be made as large as desired.  Then, in order to create a paradox, we take several of these systems in the setup of \cite{Pirani:1970bp}, all boosted relative to each other, and are capable of sending signals backwards in time to a specified location.

\subsection{Regime of Validity}
Let us carefully examine the wave solution we have constructed, to see under what conditions it is capable of giving rise to time travel paradoxes.  We restrict our attention to the spin 1 case here because it is simplest, but our conclusions are identical for all three cases.  The speed of sound in the spin 2 background is
\beq
v\sim1\mp\frac{\ddot h}{\Lambda^3}\sim1\pm\frac{\omega^2 h}{\Lambda^3}
\eeq
where we have taken the spin two laser to have a pure frequency $\omega$.  When we propagate a short wavelength test particle through this background for a distance $L$, a net buildup of time advance can be attained $\Delta L=\Delta v L$.  The bare minimum required for this to be useful for constructing a time machine is that the shift in distance be greater than the wavelength of the photon used, $\Delta L>\bar\omega^{-1}$ \cite{Hollowood:2015elj,Goon:2016une}.  At first this can seem to be made arbitrarily large, since the distance $L$ could in principle be taken to infinity.  However, we are stymied by this approach due to the fact that the wave pulse we have chosen oscillates around zero, giving the effect of a time advance half the time and a time delay the other half.  If a photon is sent through a number of cycles the effect averages to zero, and so the effect must be confined within a single wavelength, $L<\omega^{-1}$.  

Therefore, the ratio of time advance to wavelength of the photon will be
\beq
\frac{\Delta L}{\bar\omega^{-1}}\sim\frac{\bar\omega\omega^2 h L}{\Lambda^3}\sim\frac{L}{\omega^{-1}}\frac{\bar\omega}{\omega}\frac{\omega^2h}{\Lambda^3}\sim\frac{L}{\omega^{-1}}\frac{h}{h_{\text{UV}}}\frac{\bar\omega}{\omega}
\eeq
The last similarity defines the quantity $h_{\text{UV}}=\Lambda^3/\omega^2$, which is the maximum value of $h$ for which we can trust the theory.  Beyond this, higher order terms in the Lagrangian of the form $R^p F^2/\Lambda^{3p}$, which will be induced radiatively by loops, all become equally important, and the effective description breaks down.  Note that since this is the quantity that sets the change in speed, the speed itself must be very close to 1, but the total effect can be made large.  Then, the first two factors must be below 1 for our theory, but the hierarchy of photon frequency to spin 2 frequency, $\bar\omega/\omega$ can be made as large as desired.  

It is interesting to see how this effect is shielded in the case where the spin 2 field is the graviton.  In general relativity, the Lagrangian contains the additional tower of operators $h^kF^2/M_p^k$ set by the Planck mass. If the graviton becomes larger than $M_p$, the effective description breaks down, corresponding in the full theory to the creation of a black hole.  Effectively, then, the upper limit on $h$ becomes $h_{\text{UV}}=\min\{\Lambda^3/\omega^2,M_p\}$, and cannot be made arbitrarily large for small $\omega$.  Similarly, there is a tower of operators in the photon sector that scale parametrically as $\alpha^k F^{2k}/m_e^{4k-4}$ and $\Box^k F^2/m_e^{2k}$, where $m_e$ is the electron mass.  Frequencies $\bar\omega$ above the electron mass are outside the regime of the effective theory, and in the full theory correspond to the ability for a high energy photon to produce an electron-positron pair through the Schwinger mechanism.  In this case the cutoff scale $\Lambda$ we have been discussing is related to these two scales through $\Lambda^3\sim m_e^2M_p$, and so the maximal ratio is 
\beq
\frac{\Delta L}{\bar\omega^{-1}}\sim\frac{\bar\omega\omega}{m_e^2}\frac{h}{M_p}
\eeq
From the arguments above this can never be larger than one while remaining in the regime of validity of the effective theory, and so any time advance must be much shorter than the wavelength of the photon used.  This is consistent with the conclusions of \cite{Hollowood:2015elj,Goon:2016une}, who find that it is impossible to construct a time machine using the quantum effects of electrodynamics in curved space.

In this case, it is the fact that this nonminimal term is generated from extra degrees of freedom running through loops that shields the theory from superluminality.  There is no limit in which the theories of the previous section can be recovered from this theory, since the limit $M_p\rightarrow\infty$, $\Lambda$ fixed entails that $m_e\rightarrow0$.  Here, the curative degrees of freedom (electrons) are explicitly part of the low energy theory, and cannot be removed from the analysis.  This is just the simplest example of how extra degrees of freedom may enter the theory to cure this pathology.  Note, however, that it requires the addition of the minimal coupling of general relativity, rendering it disconnected from the theories we were discussing before.  Those theories, as they have been written, cannot be cured without this standard coupling included to bound the amplitude of the spin 2 wave.

\section{Two Spin 2 in the Interaction}\label{Multiple}

We now study the interactions involving several spin 2 particles, focussing on the case of two spin 2 particles at each vertex, and show that these lead to acausality as well.  The Feynman diagram for the types of interactions we are considering is depicted in Fig.~\ref{tree2} below.

\begin{centering}
\begin{figure*}[h]
\centering
\includegraphics[width=4cm]{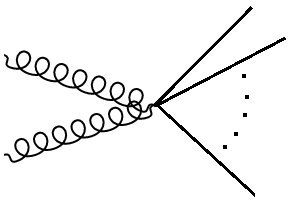}
\caption{A diagram with two spin 2 particles and an arbitrary number of matter particles, of the type studied in this section.}
\label{tree2}
\end{figure*}
\end{centering}

\subsection{Gauss-Bonnet Coupling}
We can also ask whether the Gauss-Bonnet term leads to superluminal propagation if we couple it to matter, along the lines of \cite{ChoquetBruhat:1988dw,Izumi:2014loa,Reall:2014pwa} (we note that without coupling to matter, the Gauss-Bonnet term is only a boundary term).  The matter sector remains luminal if we only use a potential function (not involving derivatives), which is also the only coupling possible without introducing ghosts (in four dimensions)\cite{Deffayet:2011gz}.  Then the theory we choose is
\beq
\cL_{\text{int}}=\sum_{\f\ff}f_{\f\ff}\left(\phi,\bar\psi\psi\right)\left(R_{\f}{}^{\mu\nu\alpha\beta}R_{\ff}{}_{\mu\nu\alpha\beta}-4R_{\f}{}^{\mu\nu}R_{\ff}{}_{\mu\nu}+R_{\f}R_{\ff}\right)
\eeq
Though we have explicitly shown that this interaction can involve multiple spin 2 fields to highlight the disconnectedness of this theory from general relativity, we now specialize to the case of a single spin 2 field for simplicity.  The equations of motion are
\beq
G^\mu_\al- R^{\mu\nu}{}_{\al\beta}\p^\beta \p_\nu f+G^\mu_\al\Box f -G^\mu_\beta \p^\beta \p_\al f-G^\beta_\al \p^\mu\p_\beta f+G^\nu_\beta\p^\beta\p_\nu f\delta^\mu_\al-\frac12R\left(\p^\mu\p_\al f-\Box f\delta^\mu_\al\right)=0
\eeq
These equations can be simplified considerably if we use an appropriate generalization of the de Donder gauge, $K^{\al\mu}(\p_\al h_{\mu\nu}-\p_\nu h_{\al\mu}/2)=0$, with $K^{\al\mu}=(\eta_{\al\mu}+\p_\al\p_\mu f)^{-1}\approx\eta^{\al\mu}-\p^\al\p^\mu f$.  These then reduce to
\beq
K^{\mu\nu}\p_\mu\p_\nu\bar h_{\al\beta}=0
\eeq
With $\bar h_{\al\beta}=h_{\al\beta}-K^{\mu\nu}h_{\mu\nu}K_{\al\beta}/2$.  If we denote $k_{||}$ and $k_{\perp}$ as the components of momentum parallel and perpendicular to $\nabla f$, the dispersion relations are then set by
\beq
(1+\ddot f)\omega^2+2\dot f' \omega k_{||}=(1-f'')k_{||}^2+k_\perp^2
\eeq
The cross term can be eliminated by using a shifted frequency variable, $\tilde\omega=\omega+\dot f'/(1+\ddot f)k_{||}$, and the dispersion relations yield the speeds of propagation
\beq
v_{||}^2=\frac{(1-f'')(1+\ddot f)+\dot f'}{(1+\ddot f)^2},\quad v_{\perp}^2=\frac{1}{1+\ddot f}
\eeq
For instance, if we take $f$ to be a function of $x+t$, then $\ddot f=f''= \dot f'$, and the sound speed in both directions becomes, to linear order, $v^2=1-\ddot f$.  Since there is no preferred sign for the second derivative of this function, the speed can be either faster or slower than luminal.

\subsection{Gravitational Chern-Simons Term}\label{GCS}

One may also wonder about a coupling between a scalar and the gravitational Chern-Simons term of the form considered in \cite{Campbell:1990fu,Jackiw:2003pm}:
\beq
\cL_{\text{int}}=\frac{1}{4}g(\phi,i\bar\psi\g_5\psi)\epsilon^{\alpha\beta\g\delta}R_{\alpha\beta\mu\nu}R^{\mu\nu}{}_{\g\delta}
\eeq
This theory has a similar status to the Gauss-Bonnet term, since it only alters the propagation of the graviton, and is a total derivative in the limit that the function $g$ becomes constant.  The equations of motion for this theory are
\beq
E^{\nu\beta}\equiv G^{\nu\beta}+\epsilon^{\mu\nu\rho\sigma}\p_\alpha\left(\p_\mu g R_{\rho\sigma}{}^{\alpha\beta}\right)+(\nu\leftrightarrow\beta)=0
\eeq

On a generic background, the function $g(\phi,i\bar\psi\g_5\psi)$ can depend on both time and space, and the equations of motion will not factorize to separate equations for each helicity.  This makes the analysis complicated, but if we make the assumption that $g$ is a small perturbation on an otherwise flat background we can still probe whether signals propagate superluminally.  The dispersion relations   can still be obtained from solutions to the equations of motion for a Fourier mode, where, since there are multiple propagating degrees of freedom, a determinant must be taken to account for all solutions.  If we abstractly represent the equations as
\beq
(A_{ab}\eta^{\mu\nu}-B^{\mu\nu}_{ab})k_\mu k_\nu=0
\eeq
To lowest order $A_{ab}=\delta_{ab}$, and the determinant can be expanded in powers of $B$.  Unfortunately this case is somewhat complicated by the fact that the trace of $B$ vanishes, but the equation can be expanded to second order to yield
\beq
(k_\mu k^\mu{})^2+\frac12\text{Tr}(B^{\mu\nu}B^{\alpha\beta})k_\mu k_\nu k_\alpha k_\beta+\mathcal{O}(g^3)=0
\eeq
and it remains to evaluate the trace of the matrix squared.  In our setting, this can be done by taking a second variation of the equations of motion, which yields
\beq
B^{\nu\beta\sigma\delta}=\frac{\delta \Delta E^{\nu\beta}}{\delta h_{\sigma\delta}}=-\epsilon^{\mu\nu\rho\sigma}\p_\alpha\p_\mu g\,k_\rho(\eta^{\delta\beta}k^\alpha-\eta^{\alpha\delta}k^{\beta})+(\nu\leftrightarrow\beta,\,\sigma\leftrightarrow\delta)+\mathcal{O}(k^3)
\eeq
Here we have ignored higher powers of the momentum, as on backgrounds with spatial dependence these will only introduce spurious poles with a mass proportional to $m\sim 1/g'$, outside the regime of validity of our theory.  This matrix must be projected onto physical degrees of freedom before the trace is taken, for which we use the flat space projector $P_{ijkl}=\hat\delta_{ik}\hat\delta_{jl}+\hat\delta_{il}\hat\delta_{jk}-\hat\delta_{ij}\hat\delta_{kl}$, with $\hat\delta_{\ij}=\delta_{ij}-\hat k_i\hat k_j$.  At this point the computation can be performed, yielding
\beq
(\omega^2-k^2)^2+\left|\omega\ddot g\vec{k}-k\cdot\nabla g\vec{k}-\omega^2\vec{\nabla}\dot g+\omega\, k\cdot\nabla\vec{\nabla}g\right|^2+\mathcal{O}(g^3)=0
\eeq
This generalizes earlier work of Ref.~\cite{Dyda:2012rj} to spatially dependent backgrounds.  The dispersion relations set by solutions to this equation will still retain their complex sound speeds, but we remain agnostic as to whether this in itself is a problem.  The pathology we are more interested in is whether there are backgrounds for which the real part of the propagation speed is greater than one, signalling superluminal motion.  We solve this quartic equation for $\omega(k)$ to second order to obtain
\beq
\text{Re}\{v\}=1-\frac18\left(\beta_1^2-4\beta_1\beta_2+\beta_2^2+2\beta_1\beta_3\right)+\mathcal{O}(g^3)
\eeq
where we have defined the angle-dependent quantities $\beta_1=\ddot g$, $\beta_2=\hat k\cdot\nabla\dot g$, $\beta_3=\hat k\cdot\nabla\hat k\cdot\nabla g$.  To explicitly show that this dispersion relation leads to superluminal motion, we specialize to a setup where we have a long wavelength mode for $g$ propagating in the $z$ direction.  Then the speed becomes
\beq
\text{Re}\{v\}=1-\frac{g''^2}{8}\left(1-4\cos\theta+3\cos^2\theta\right)
\eeq
which is indeed greater than 1 for angles between $0<\cos\theta<1/3$.  Thus on spatially dependent backgrounds the gravitational Chern-Simons theory can give rise to superluminal propagation.

\subsection{Quantum Corrections for Fermions}\label{fermRR}

\begin{centering}
\begin{figure*}[h]
\centering
\includegraphics[width=6cm]{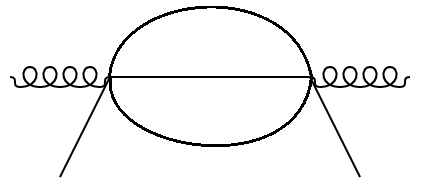}
\caption{While the theory given by equation \ref{Rpsi4} is not itself unhealthy, pathological terms are generated through the above two loop diagram, leading it to induce theories of the type studied in section \ref{GCS}.}
\label{2ool}
\end{figure*}
\end{centering}

We now discuss that the seemingly healthy term in equation \ref{Rpsi4} generates the Chern-Simons term through loop effects.  The loop under discussion is depicted in Fig.~\ref{2ool}. 
This term involves four fermions and a single spin 2 field and was a dimension 9 operator.  It will lead to the Chern-Simons operator involving two fermions and two spin 2 fields, which is also dimension 9, and so can potentially play the leading role for the dynamics of the theory.  It is generated through two loop diagrams, but we refrain from going into too much detail about the calculation, as it is quite beside the main topic of this paper.  Suffice to say, the effect can be encapsulated with the following terms in the effective Lagrangian:
\beq
\cL_{\text{eff}}=\frac{1}{\Lambda^{10}}R_{\alpha\beta\gamma\delta}R_{\mu\nu\rho\sigma}\left(m^3I_2+mI_4\right)\bar\psi\left(J_1^{\alpha\beta\gamma\delta\mu\nu\rho\sigma}+J_2^{\alpha\beta\gamma\delta\mu\nu\rho\sigma}\right)\psi
\eeq
where $I_2$ and $I_4$ are quadratically and quartically divergent integrals, respectively, $m$ is the mass of the fermion running through the loops, and $J_1$ and $J_2$ involve the index structure of the $\sigma$ matrices coming from the two possible Wick contractions of the two loop diagram.  Explicitly, we have $J_1^{\alpha\beta\gamma\delta\mu\nu\rho\sigma}=\sigma^{\alpha\beta}\sigma^{\mu\nu}\sigma^{\g\delta}\sigma^{\rho\sigma}$ and $J_2^{\alpha\beta\gamma\delta\mu\nu\rho\sigma}=\sigma^{\alpha\beta}\sigma^{\rho\sigma}\text{tr}(\sigma^{\g\delta}\sigma^{\mu\nu})$.  Where these multiply the quartic divergence, the sigma matrices are accompanied by another pair of gamma matrices contracted among themselves, inserted at appropriate locations.

Using the gamma matrix identities $[\sigma^{\mu\nu},\sigma^{\gamma\delta}]=2i\left(\eta^{\mu\gamma}\sigma^{\nu\delta}-\eta^{\nu\gamma}\sigma^{\mu\delta}-\eta^{\mu\delta}\sigma^{\nu\g}+\eta^{\nu\delta}\sigma^{\mu\g}\right)$ and equation \ref{anticomm} for the anticommutator, the sigma matrices in these expressions can be eliminated in favor of $\eta^{\mu\nu}$ and $\epsilon^{\alpha\beta\gamma\delta}\g_5$.  In all cases two commutators and one anticommutator are used, ultimately introducing only a single $\epsilon$ tensor, which contracts to form the Chern-Simons combination.

The fact that only the Chern-Simons combination is generated, as opposed to the Gauss-Bonnet of the same dimension, can be seen because only this term retains the symmetry of the original term $\psi_R\rightarrow i\psi_L$, $\psi_L\rightarrow i\psi_R$.  The actual coefficient of the Chern-Simons term is ambiguous in the effective field theory framework since it is proportional to a power-law divergent integral, but in the absence of any motivation otherwise we take the ultimate coefficient to be the same scale as the original term in the Lagrangian, $\Lambda^{-5}$.  In this case the original term will not affect the causality properties of the theory, but the term it generates through loops will induce superluminality.

\section{General Analysis}\label{General}

The interactions we have presented to this point are by no means an exhaustive list.  There is no limit to the number of operators one may write down that represent a nonminimal coupling of a spin 2 field to other matter representations.  For instance, one may contemplate all possible quartic interactions between two spin 2 fields and two spin 1 fields.  These will all have ghosts in four dimensions, as their equations of motion will be third order.  If one is willing to consider six dimensions, however, a combination employing the Levi-Civita symbols can be written, which is ghost free.
Similarly, in \cite{Camanho:2014apa} cubic self interactions of spin 2 particles were analyzed, such as
$\Delta \cL \propto R_{\mu\nu\sigma\delta}R^{\sigma\delta\rho\gamma}R_{\rho\gamma}^{\mu\nu}+\ldots$,
where it was shown that it generically gives rise to superluminality.  Faced with this never ending panoply of possible couplings, we are motivated to provide a more general analysis that lays out the conditions that will lead to superluminality.

\subsection{Conditions for Superluminality}

We now formulate a statement that a generic coupling of matter to the Riemann tensor induces superluminality on some background.  This is necessary, as otherwise there are in principle an infinite number of couplings that could be chosen, and each would potentially represent a healthy theory.  This proof eliminates that worry.

In all cases we consider, the equations of motion for the test particle in the slowly varying spin 2 background reduces to
\beq
(\eta^{\mu\nu}-B^{\mu\nu}(h))k_\mu k_\nu=0\label{generalB}
\eeq
This form is essential for the consistency of a theory.  While generic couplings would yield equations of motion with higher powers of momenta, these would all have ghostlike roots.  If the mass of the ghost is field dependent it can be made arbitrarily small for some backgrounds, and the theory would have a vanishing regime of validity.  Alternatively, if the mass of the ghost is heavy we ignore those branches as artefacts of the effective field theory description, and focus on the quadraticized version of the equations of motion describing the healthy modes.\footnote{The one exception to this reasoning would be if the extra terms were only spatial, so that additional roots to the dispersion relations are not introduced.  This can only occur for a single Lorentz frame, but nevertheless if the background breaks Lorentz invariance then it may occur.  In this case there are three possible behaviors:  the speed can either become imaginary, superluminal, or neither for large momenta.  The first would arise in dispersion relations of the form $\omega^2=k^2-\alpha k^4$, which exhibits gradient instabilities for large $k$.  The second is present for theories like $\omega^2=k^2+\alpha k^4$, which are ruled out by superluminality considerations.  The last are present for theories like $\omega^2=k\tanh(\alpha k)$, such as water waves.  These types of theories necessarily require an infinite number of operators to give such a result, and so are not of the form we consider in effective field theory.  Let us note that the $R^4$ theories that arise in low energy descriptions of string theories are of this last type \cite{Gruzinov:2006ie,Bellazzini:2015cra}: subluminal, but with spurious poles that can only be removed by an infinite tower of operators coming from a more complete theory.}

Granted that the theory we consider prepares this dispersion relation, our requirement that it be subluminal reduces to $\text{Re}\{v\}<1$, where $\omega^2=v^2|k|^2$ and the speed may depend on the particular direction of travel.  Note that in these theories we need not make the distinction between group velocity, phase velocity, and signal velocity, since the dispersion relation is exactly quadratic.  The speed can be found from this equation as
\beq
-(1+B^{00})\omega^2-2B^{0i}\omega k_i+(\delta^{ij}-B^{ij})k_ik_j=0
\eeq
or, in terms of the shifted frequency $\tilde\omega=\omega+B^{0i}k_i/(1+B^{00})$,
\beq
\tilde\omega^2=\frac{1}{1+B^{00}}\left(\delta^{ij}-B^{ij}+\frac{B^{0i}B^{0j}}{1+B^{00}}\right)k_ik_j
\eeq
which gives the propagation speed as
\beq
v^2=\frac{1}{1+B^{00}}\left(1-B^{ij}+\frac{B^{0i}B^{0j}}{1+B^{00}}\right)\hat k_i\hat k_j
\eeq
If we take the matrix $B$ to be a perturbation we can ignore the last term, as it is higher order.  Then the sound speed attains a maximum at
\beq
v^2_{max}=\frac{1-b_{\text{min}}}{1+B^{00}}\label{cmax}
\eeq
where $b_{\text{min}}$ is the smallest eigenvalue of $B^{ij}$.  Requiring that this be no greater than one implies that $B^{00}+b_{\text{min}}\geq0$.  This is equivalent to a null energy condition for the matrix $B_{\mu\nu}$.  

Including the time-space components only increases the speed, so that if this equation is not satisfied for the regime it can be neglected, then it will only get worse for the regime where it cannot.  In fact, this stipulation can be avoided by noting that the null energy condition is independent of Lorentz frame, so that if we can find a frame in which the time-space components vanish, the equivalence is exact.  However, the conditions for such a frame to exist are precisely the null energy condition, leading us to the general statement that 
\beq
v^2<1\iff B^{\mu\nu}k_\mu k_\nu\geq0\,\, \forall\,\, k^2=0\label{nec}
\eeq

But from here it becomes easy to see that theories of the type we consider will always lead to superluminalities, because they are linear in the spin 2 field.  So, if we find one wave solution that has subluminal propagation, the solution with $R_{\mu\nu\al\beta}\rightarrow-R_{\mu\nu\al\beta}$ will have superluminal propagation.  Since there is nothing in the theory that sets a preferred sign for the spin 2 field, generically faster than light travel will exist in the matter sector.  This holds for arbitrary Lagrangians of the form $\Delta\cL=R_{\mu\nu\alpha\beta}T^{\mu\nu\alpha\beta}$, where $T$ contains derivative terms.  However, one may also be concerned with interactions involving two powers of the Riemann curvature, and so on.  These will lead to superluminalities for the exact same reason if we consider backgrounds with a superposition of multiple spin 2 modes.  For instance, if we take one to be extremely large wavelength so that it is effectively a constant background for the scales we are interested in, we can set up a scenario on top of this that will operate in exactly the same manner as above.  We need not worry about any nonlinearities in the spin 2 field equations because we are free to take the frequency of the long mode to be arbitrarily small, and compensate this suppression by increasing the amplitude of the shorter wavelength mode.

Finally, we briefly comment on the generalization of \ref{nec} to the case of multiple species, as we needed for our analysis in section \ref{GCS}.  In this case $B^{\mu\nu}$ will be a matrix, and the determinant of equation \ref{generalB} will yield roots corresponding to all dispersion relations of the theory.  If $B$ is treated as a perturbation this can be expanded to first order, and the condition for superluminality is simply $\text{Tr}B^{\mu\nu}k_\mu k_\nu\geq0$ for null $k_\mu$.  Should the scenario arise where the trace of this matrix vanishes, as in the Chern-Simons case discussed above, then the condition is $\text{Tr}(B^{00}B^{\mu\nu})k_\mu k_\nu\geq0$.

\section{Comparison to Other Theories}\label{Comparison}

\subsection{General Relativity}

It is worth asking how general relativity manages to always satisfy this constraint, since gravity modifies the propagation of light rays in exactly this way, yet retains a notion of causality.  We consider two scenarios, that when the metric is set by a weak field source, and the case of a gravitational wave.  

If the Einstein tensor is sourced by matter, then, following \cite{Visser:1998ua}, the induced nondynamical graviton profile is
\beq
h^{\mu\nu}=\frac{\kappa}{2}\int d^4y\,G_{\text{ret}}(x,y)\left(T^{\mu\nu}(y)-\frac12T^\lambda{}_\lambda(y)\eta^{\mu\nu}\right)
\eeq
where we have expanded around flat space in Lorentz gauge, and used the retarded solution to the wave equation that has support only within the past light cone.  Then $B^{\mu\nu}=h^{\mu\nu}$ and from (\ref{cmax}) the speed is given by
\beq
v^2=\frac{1-\langle T_{\text{min}}\rangle+\frac12\langle T^\mu{}_\mu\rangle}{1+\langle T^{00}\rangle+\frac12\langle T^\mu{}_\mu\rangle}
\eeq
where $T_{\text{min}}$ is the minimum eigenvalue of the spatial part of the stress tensor and we introduced the bracket notation to denote convolution with the retarded Green's function $G_{\text{ret}}$.  Keeping in mind that the Green's function is positive semidefinite, the condition for this to be subluminal is $T^{00}+T_{\text{min}}\geq0$, which is the null energy condition.  Using the additional stipulation that the energy is positive, this shows that our criterion reproduces the standard result \cite{Tipler:1977eb,Olum:1998mu} that the weak energy condition must hold to avoid superluminal motion.

Let us draw to the reader's attention that this condition is a requirement for causality in the purely classical theory.  This will not be satisfied when treated as a quantum field theory, in generic quantum states.  However, this does not immediately imply a violation of causality, as any amount of negative energy must necessarily be localized to a small region of spacetime \cite{fordroman}.  The requirement for causality is then replaced with the averaged weak energy condition \cite{worm}, in which the contribution from every point on the particle's path is integrated over.  While this stricter criterion is technically the requirement for acausality, it is of no interest in the nonminimally coupled cases, where causality is violated even in the classical theory, not to mention the quantum version.

The case for a gravitational wave is equally instructive.  Here we are free to take the transverse traceless gauge\footnote{This coordinate system will not be geodesically complete in the interacting theory \cite{Penrose:1965rx}, which we could use as a prompt to go to the pp spacetime gauge.  However, we are free to choose a setup that staves off this incompleteness to arbitrarily far distances, and so this will not affect our analysis.}, in which the dispersion relation for a test particle becomes
\beq
\omega^2=(\delta_{ij}+h_{ij})k_ik_j
\eeq
The speeds can be read off as $v^2=\{1,1\pm\sqrt{h_+^2+h_\times^2}\}$.  From this it appears that one of the directions does in fact become superluminal.  However, one should be careful in this case, because there are number of subtleties.  First of all, because of the equivalence principle, the term superluminal is a misnomer, as now all null curves will be affected in exactly the same manner.  Secondly, in diffeomorphism invariant theories we need to make sure this does not just represent a rescaling of the coordinates, either globally or locally.  However, the presence of a graviton is not something that can be gauged away with a coordinate transformation, and so represents a physical effect.  What is nontrivial is the fact that even though passing through a graviton induces superluminality, it cannot lead to a time machine in general relativity because gravity is necessarily self interacting due to the equivalence principle.  Then, if one tries to engineer a setup where a photon is passed through a boosted graviton and sent back through another boosted graviton to send future-directed signals back in time, one finds that the presence of the first graviton automatically induces an equal shift in the location of the second to completely cancel the net effect, as found in \cite{Shore:2002in,Hollowood:2015elj}.  Thus, causality is preserved in general relativity. On the other hand, acausality can be present in the much larger class of derivative theories examined earlier, because there the spin 2 field does not interact with itself (at the classical level), and so two waves can pass by each other by superposition.

\subsection{Spin 1}\label{spinny1}

We briefly consider nonminimal couplings of a spin 1 field to matter to highlight some of the differences between this and the spin 2 case.  If we take the coupling to scalars, the couplings with the lowest number of derivatives are
\beq
\cL_{\text{int}}=f(\phi)F_{\mu\nu}F^{\mu\nu}+g(\phi)\epsilon^{\mu\nu\alpha\beta}F_{\mu\nu}F_{\alpha\beta}
\eeq
These will not alter the causal structure of the scalar, but will affect the spin 1 field.  The first `dilaton type' coupling only serves to multiply the eikonal equation by $(1+f(\phi))$.  As long as this is positive, the theory is healthy.  The second `axion type' coupling is a bit more subtle, since on a time dependent $\phi$ background the dispersion relation for the spin 1 field becomes \cite{Carroll:1989vb}
\beq
\left(1+f\right)\left(\omega^2-k^2\right)=\pm\dot g k
\eeq
This term violates parity, but, in contrast to the gravitational Chern-Simons case we discussed in section \ref{GCS}, the effects only occur at \emph{low} energies as opposed to high energies.  Thus, one of the modes will become a tachyon for $k<\dot g$, and the other will appear superluminal.  But this apparent superluminality is not real, as the high momentum limit recovers the Lorentz invariant dispersion relation, guaranteeing that signals propagate on the light cone.

An analog for the Riemann coupling to fermions can be found in the case of electromagnetism.  We are free to write down a term of the form
\ba
&&\cL_{\text{int},1}=\frac{1}{\Lambda}\bar\psi(e_1 F^{\mu\nu}+ie_2\tilde F^{\mu\nu}\g_5)\sigma_{\mu\nu}\psi\\
&&\cL_{\text{int},2}=\frac{i}{\Lambda^2}\bar\psi(e_3 F^{\mu\nu}+ie_4\tilde F^{\mu\nu}\g_5)\g_\mu\p_\nu\psi\label{anapolis}
\ea
The first $\cL_{\text{int},1}$ are dipole moment interactions and the second $\cL_{\text{int},2}$ are anapole moment interactions as discussed in \cite{Nowakowski:2004cv,Giunti:2008ve,Broggini:2012df,Giunti:2015gga}.  As far as its derivative structure is concerned, the first is a kind of complicated mass term for the fermion in the presence of a background electromagnetic field, and hence is causal.

The second will alter the kinetic term to leading order.  We have written down a general combination of the field strength and its dual.  The limit $e_3=0$ reproduces the term responsible for the Velo-Zwanziger pathology \cite{Velo:1969bt}.  For a massive charged spin $3/2$ minimally coupled to electromagnetism, it is well known that in the presence of a background magnetic field, the longitudinal mode exhibits superluminal propagation\cite{MadoreTait,Madore,Deser:2001dt}.  If one introduces the Stueckelberg mode and takes the decoupling limit of the Lagrangian, a term exactly of the form of (\ref{anapolis}) is present \cite{Rahman:2011ik}.  There it was shown to be removable with a field redefinition, at the expense of introducing fermion self couplings.  So though it exhibits superluminality, this does not represent a real interaction with the spin 1 particle.

\subsection{Massive Spin 2}

In previous sections, our analysis focused on the case of massless spin 2 particles, but it is not hard to generalize our results to the case of massive particles.  Generically, these theories were shown to exhibit faster than light propagation \cite{Deser:2012qx,Deser:2014hga}, and this behavior can be expected to hold in the case of nontrivial coupling as well.  We will show that this expectation is indeed borne out.

Generically, a mass term can be expected to break the symmetry that eliminates the lower spin components from the kinetic term.  These will acquire dynamics, as is most easily seen by employing the Stueckelberg trick of restoring the gauge symmetry by explicitly adding fields representing the longitudinal modes into the Lagrangian description.  Since the purely longitudinal mode contains two derivatives, a generic mass term for the spin two field will lead to equations of motion that are fourth order in derivatives.  The only combination capable of evading this is the Fierz-Pauli mass, $\Delta\cL=\frac12m^2(h^2-h_{\mu\nu}h^{\mu\nu})$, where the potentially dangerous terms assemble into a total derivative.  

Let us now consider coupling this massive field to matter. We can choose any of the interactions discussed earlier which make use of the linearized Riemann tensor. These all lead to superluminality in the same way for the massless case; so we will not repeat that analysis here. 

Moreoever, there are new types of interactions allowed for massive spin 2 that make use of the field $h_{\mu\nu}$ itself. We focus on coupling to scalar matter for simplicity.  The simplest possible nonminimal coupling is
\beq
\Delta\cL=\frac12f(\phi)\left(h^2-h_{\mu\nu}h^{\mu\nu}\right)
\eeq
This resembles a part of quasi-dilaton massive gravity \cite{D'Amico:2012zv}, where it was showed that this coupling is ghost free.  Isolating the longitudinal mode by performing the usual Stueckelberg replacement 
\beq
h_{\mu\nu}\rightarrow h_{\mu\nu}+\frac{1}{\sqrt{3}}\left(\frac1f\partial_\mu\partial_\nu\pi+\pi\eta_{\mu\nu}\right),
\eeq
we find that the longitudinal mode now propagates according to 
\beq
(\eta^{\mu\nu}+\frac13\Box f^{-1}\eta^{\mu\nu}-\frac13\p^\mu\p^\nu f^{-1})\p_\mu\p_\nu\pi=0
\eeq
This is of the form (\ref{generalB}), and so gives rise to superluminal motion on some backgrounds.

The most familiar possibility is to minimally coupling $h_{\mu\nu}$ to matter, which takes the form $ \Delta\cL=-\kappa\,h_{\mu\nu}T^{\mu\nu}_{\text{M}}/2$ at leading order in the coupling $\kappa$. The need to couple to the matter energy momentum here arises from the need to avoid the 6th degree of freedom, which is a ghost. However, as is well known, this theory is only ghost free at $\mathcal{O}(\kappa)$ and requires an infinite tower of corrections to build a truly ghost free theory including self interactions; as laid out in Ref.~\cite{deRham:2010kj}. However such theories involve non-trivial self interactions of the scalar mode of the graviton of the Galileon form $\propto\Box\pi(\partial\pi)^2$, which are again associated with superluminality.

\section{Conclusions}\label{Conclusions}

In summary, we find that all the leading order couplings of spin 2 particles to matter are acausal, except in the special case of general relativity. This appears then to be the universal principle that selects general relativity as the correct theory of spin 2 at low energies, and in particular selects the number of species of spin 2 to be just one. 

This extends the results of previous uniqueness proofs that relied on the leading interaction being that of a minimally coupled field, strengthening this to more general momentum dependence of the interaction.  Since causality is usually regarded as necessary for a theory to avoid pathologies and is also closely related to unitarity in a quantum field theory, no additional input is needed to derive general relativity's primacy among the possible space of spin 2 theories.

Though the style of our analysis was to analyze example theories one by one and explicitly show that they all contain this pathology, the subtleties we encountered in this exercise allowed us to formalize the conditions for any theory to be superluminal: as long as the dispersion relations reduce to a quadratic function of the four-momentum and there is some background for which the corrections to the dispersion relation can have either sign,
the theory will be superluminal.  

In principle there could be causal sub-leading interactions at sufficiently high mass dimension which violate the conditions for our no-go theorem. An example is to take some of the operators mentioned here and squaring them, such as $\sim(RF^2)^2$ or $\sim(R_IR_J)^2$, and assigning positive coefficients in the Lagrangian (though this tends to introduces ghosts). 
But our focus here has primarily been on the most leading order interactions, which ordinarily dominate at low energies. 

We note that in the case of general relativity, plus non-minimal corrections, the situation with regards to causality is fundamentally different. In this case, the leading effects at low energies is provided by the usual general relativistic time delay for any ordinary source (Shapiro time delay). Then the non-minimal coupling adds a secondary correction that can involve time advance, which is irrelevant at low energies and is more relevant at higher energies. Then the scale at which these two effects balance each other defines a cutoff on the effective field theory, because above this scale there would be acausality; this demands new physics at or below this scale.

However, in this much larger class of theories we are studying with multiple species of spin 2, which prevents the inclusion of minimal coupling (or a single species of spin 2 with $G_N = 0$), these non-minimal effects are the leading effects. So we immediately have acausality even at arbitrarily low energies. This cannot be fixed by introducing new physics at some finite energy scale. Hence such theories cannot be UV completed in any conventional Wilsonian sense; it strongly suggests a fundamentally pathological theory.

It is striking that of the entire possible space of spin 2 theories studied, only a single one appears internally consistent, and represents the physics that we observe in our world.  This, along with the prior results that causality can restrict derivative self interactions \cite{Adams:2006sv,Cheung:2016drk}, establishes that causality can be an extremely powerful constraint. 

Also striking is the stark contrast between these considerations for spin 1 theories: generic nonminimal couplings between spin 1 fields and matter are healthy.  This makes the spin 2 case very special, and leaves general relativity as especially unique.

\section*{Acknowledgments}
We would like to thank the Tufts Institute of Cosmology for support. This work is supported by National Science Foundation grant PHY-1720332.

\section{Appendix: Field Redefinitions and Decoupling}\label{Appendix}
We will prove here that any interaction among the spin 2 field and matter proportional to the Einstein tensor can be removed with a field redefinition, so that the theory becomes that of a free spin two field and a more complicated matter sector.  
Any theory of this form can be cast as 
\beq
\cL=h^{\mu\nu}G_{\mu\nu}+G^{\mu\nu}S_{\mu\nu}
\eeq
where $S_{\mu\nu}$ is a general function of the matter fields, and we have integrated the spin 2 kinetic term by parts.  For example, in the scalar Lagrangian (\ref{scal}) we consider first, we would write 
\beq
S_{\mu\nu}\equiv-\left(a_1f(\phi)+a_2\p\phi^2\right)\eta_{\mu\nu}+a_3\p_\mu\phi\p_\nu\phi
\eeq
To track how the Einstein tensor changes under field redefinition, we use the expression
\beq
G^\mu_\alpha=\epsilon^{\mu\nu\rho\sigma}\epsilon_{\alpha\nu\beta\gamma}\p_\rho\p^\beta h^\gamma_\sigma
\eeq
This form makes it manifest that this tensor is conserved, and reproduces the kinetic term (\ref{kin}), up to boundary terms.  The Lagrangian can be rewritten as
\beq
\cL=\epsilon^{\al\beta\g\delta}\epsilon_{\al\mu\nu\rho}\left(h^\mu_\beta+S^\mu_\beta\right)\p_\g\p^\nu h^\rho_\delta
\eeq
Now, if we redefine the spin 2 field as
\beq
h_{\mu\nu}\rightarrow\tilde h_{\mu\nu}=h_{\mu\nu}-\frac12S_{\mu\nu},
\eeq
we affect a decoupling of the two sectors.  Then, the Lagrangian is recast as  
\ba
\cL&=&\epsilon^{\al\beta\g\delta}\epsilon_{\al\mu\nu\rho}\left(\tilde h^\mu_\beta\p_\g\p^\nu \tilde h^\rho_\delta-\frac14 S^\mu_\beta\p_\g\p^\nu S^\rho_\delta\right)\nn\\
&=&\tilde h^{\mu\nu}\tilde G_{\mu\nu}+\frac18\left(S^{\beta\delta}(\Box S_{\beta\delta}-2\p_\beta\p^\g S_{\g\delta})+S(2\p^\mu\p^\nu S_{\mu\nu}-\Box S)\right)
\ea

Hence, instead of there being an interaction between the spin 2 field and matter, which could potentially be causal, there is in fact only self interactions in the matter sector. Some self interactions of this nature have been studied in previous works \cite{Adams:2006sv}, and these results have recently been systematically extended in \cite{Cheung:2016drk}.  This holds completely generally, even with multiple species of matter or spin 2 fields, and allows us to focus our attention on couplings to the Riemann tensor, which represent true interactions that cannot be removed in this way.

Let us contrast this procedure to the one done ubiquitously in gravitational theories: because the kinetic term is quadratic and the interaction term is linear, this decoupling is \emph{exact}.  In gravitational theories this is not the case because the theory is nonlinear, and the field redefinition we have given only serves to stave off interactions to higher orders in the fields.


\newpage
\begin{thebibliography}{99}
\bibitem{Hertzberg:2016djj} 
  M.~P.~Hertzberg,
  ``Gravitation, Causality, and Quantum Consistency,''
  arXiv:1610.03065 [hep-th].


\bibitem{Wald:1986bj} 
  R.~M.~Wald,
  ``Spin-2 Fields and General Covariance,''
  Phys.\ Rev.\ D {\bf 33}, 3613 (1986).
  doi:10.1103/PhysRevD.33.3613


\bibitem{Bai:2016hui} 
  D.~Bai and Y.~H.~Xing,
  ``Special Gravity as Alternatives for Interacting Massless Gravitons,''
  arXiv:1610.00241 [hep-th].


\bibitem{Bai:2017dwf} 
  D.~Bai and Y.~H.~Xing,
  ``On the uniqueness of ghost-free special gravity,''
  arXiv:1702.05756 [hep-th].

\bibitem{Boulanger:2000rq} 
  N.~Boulanger, T.~Damour, L.~Gualtieri and M.~Henneaux,
  ``Inconsistency of interacting, multigraviton theories,''
  Nucl.\ Phys.\ B {\bf 597}, 127 (2001)
  doi:10.1016/S0550-3213(00)00718-5
  [hep-th/0007220].

\bibitem{Benincasa:2007xk} 
  P.~Benincasa and F.~Cachazo,
  ``Consistency Conditions on the S-Matrix of Massless Particles,''
  arXiv:0705.4305 [hep-th].
  
\bibitem{Deser:1969wk} 
  S.~Deser,
  ``Selfinteraction and gauge invariance,''
  Gen.\ Rel.\ Grav.\  {\bf 1}, 9 (1970)
  doi:10.1007/BF00759198
  [gr-qc/0411023].


\bibitem{Deffayet:2011gz} 
  C.~Deffayet, X.~Gao, D.~A.~Steer and G.~Zahariade,
  ``From k-essence to generalised Galileons,''
  Phys.\ Rev.\ D {\bf 84}, 064039 (2011)
  doi:10.1103/PhysRevD.84.064039
  [arXiv:1103.3260 [hep-th]].


\bibitem{Ohashi:2015fma} 
  S.~Ohashi, N.~Tanahashi, T.~Kobayashi and M.~Yamaguchi,
  ``The most general second-order field equations of bi-scalar-tensor theory in four dimensions,''
  JHEP {\bf 1507}, 008 (2015)
  doi:10.1007/JHEP07(2015)008
  [arXiv:1505.06029 [gr-qc]].


\bibitem{Ohkuwa:1980jx} 
  Y.~Ohkuwa,
  ``Effect of a Background Gravitational Field on the Velocity of Neutrinos,''
  Prog.\ Theor.\ Phys.\  {\bf 65}, 1058 (1981).
  doi:10.1143/PTP.65.1058


\bibitem{Drummond:1979pp} 
  I.~T.~Drummond and S.~J.~Hathrell,
  ``QED Vacuum Polarization in a Background Gravitational Field and Its Effect on the Velocity of Photons,''
  Phys.\ Rev.\ D {\bf 22}, 343 (1980).
  doi:10.1103/PhysRevD.22.343


\bibitem{Shore:2003zc} 
  G.~M.~Shore,
  ``Quantum gravitational optics,''
  Contemp.\ Phys.\  {\bf 44}, 503 (2003)
  doi:10.1080/00107510310001617106
  [gr-qc/0304059].


\bibitem{Pirani:1970bp} 
  F.~A.~E.~Pirani,
  ``Noncausal behavior of classical tachyons,''
  Phys.\ Rev.\ D {\bf 1}, 3224 (1970).
  doi:10.1103/PhysRevD.1.3224


\bibitem{Hollowood:2015elj} 
  T.~J.~Hollowood and G.~M.~Shore,
  ``Causality Violation, Gravitational Shockwaves and UV Completion,''
  JHEP {\bf 1603}, 129 (2016)
  doi:10.1007/JHEP03(2016)129
  [arXiv:1512.04952 [hep-th]].


\bibitem{Goon:2016une} 
  G.~Goon and K.~Hinterbichler,
  ``Superluminality, Black Holes and Effective Field Theory,''
  arXiv:1609.00723 [hep-th].


\bibitem{ChoquetBruhat:1988dw} 
  Y.~Choquet-Bruhat,
  ``The Cauchy Problem for Stringy Gravity,''
  J.\ Math.\ Phys.\  {\bf 29}, 1891 (1988).
  doi:10.1063/1.527841


\bibitem{Izumi:2014loa} 
  K.~Izumi,
  ``Causal Structures in Gauss-Bonnet gravity,''
  Phys.\ Rev.\ D {\bf 90}, no. 4, 044037 (2014)
  doi:10.1103/PhysRevD.90.044037
  [arXiv:1406.0677 [gr-qc]].


\bibitem{Reall:2014pwa} 
  H.~Reall, N.~Tanahashi and B.~Way,
  ``Causality and Hyperbolicity of Lovelock Theories,''
  Class.\ Quant.\ Grav.\  {\bf 31}, 205005 (2014)
  doi:10.1088/0264-9381/31/20/205005
  [arXiv:1406.3379 [hep-th]].


\bibitem{Campbell:1990fu} 
  B.~A.~Campbell, M.~J.~Duncan, N.~Kaloper and K.~A.~Olive,
  ``Gravitational dynamics with Lorentz Chern-Simons terms,''
  Nucl.\ Phys.\ B {\bf 351}, 778 (1991).
  doi:10.1016/S0550-3213(05)80045-8


\bibitem{Jackiw:2003pm} 
  R.~Jackiw and S.~Y.~Pi,
  ``Chern-Simons modification of general relativity,''
  Phys.\ Rev.\ D {\bf 68}, 104012 (2003)
  doi:10.1103/PhysRevD.68.104012
  [gr-qc/0308071].


\bibitem{Dyda:2012rj} 
  S.~Dyda, E.~E.~Flanagan and M.~Kamionkowski,
  ``Vacuum Instability in Chern-Simons Gravity,''
  Phys.\ Rev.\ D {\bf 86}, 124031 (2012)
  doi:10.1103/PhysRevD.86.124031
  [arXiv:1208.4871 [gr-qc]].


\bibitem{Camanho:2014apa} 
  X.~O.~Camanho, J.~D.~Edelstein, J.~Maldacena and A.~Zhiboedov,
  ``Causality Constraints on Corrections to the Graviton Three-Point Coupling,''
  JHEP {\bf 1602}, 020 (2016)
  doi:10.1007/JHEP02(2016)020
  [arXiv:1407.5597 [hep-th]].


\bibitem{Gruzinov:2006ie} 
  A.~Gruzinov and M.~Kleban,
  ``Causality Constrains Higher Curvature Corrections to Gravity,''
  Class.\ Quant.\ Grav.\  {\bf 24}, 3521 (2007)
  doi:10.1088/0264-9381/24/13/N02
  [hep-th/0612015].


\bibitem{Bellazzini:2015cra} 
  B.~Bellazzini, C.~Cheung and G.~N.~Remmen,
  ``Quantum Gravity Constraints from Unitarity and Analyticity,''
  Phys.\ Rev.\ D {\bf 93}, no. 6, 064076 (2016)
  doi:10.1103/PhysRevD.93.064076
  [arXiv:1509.00851 [hep-th]].


\bibitem{Visser:1998ua} 
  M.~Visser, B.~Bassett and S.~Liberati,
  ``Superluminal censorship,''
  Nucl.\ Phys.\ Proc.\ Suppl.\  {\bf 88}, 267 (2000)
  doi:10.1016/S0920-5632(00)00782-9
  [gr-qc/9810026].


\bibitem{Tipler:1977eb} 
  F.~J.~Tipler,
  ``Singularities and Causality Violation,''
  Annals Phys.\  {\bf 108}, 1 (1977).
  doi:10.1016/0003-4916(77)90348-7


\bibitem{Olum:1998mu} 
  K.~D.~Olum,
  ``Superluminal travel requires negative energies,''
  Phys.\ Rev.\ Lett.\  {\bf 81}, 3567 (1998)
  doi:10.1103/PhysRevLett.81.3567
  [gr-qc/9805003].

  \bibitem{fordroman} 
  L.~H.~Ford and T.~A.~Roman,
  ``Averaged energy conditions and quantum inequalities,''
  Phys.\ Rev.\ D {\bf 51}, 4277 (1995)
  doi:10.1103/PhysRevD.51.4277
  [gr-qc/9410043].

\bibitem{worm} 
  M.~S.~Morris, K.~S.~Thorne and U.~Yurtsever,
  ``Wormholes, Time Machines, and the Weak Energy Condition,''
  Phys.\ Rev.\ Lett.\  {\bf 61}, 1446 (1988).
  doi:10.1103/PhysRevLett.61.1446
  

\bibitem{Penrose:1965rx} 
  R.~Penrose,
  ``A Remarkable property of plane waves in general relativity,''
  Rev.\ Mod.\ Phys.\  {\bf 37}, 215 (1965).
  doi:10.1103/RevModPhys.37.215


\bibitem{Shore:2002in} 
  G.~M.~Shore,
  ``Constructing time machines,''
  Int.\ J.\ Mod.\ Phys.\ A {\bf 18}, 4169 (2003)
  doi:10.1142/S0217751X03015118
  [gr-qc/0210048].


\bibitem{Carroll:1989vb} 
  S.~M.~Carroll, G.~B.~Field and R.~Jackiw,
  ``Limits on a Lorentz and Parity Violating Modification of Electrodynamics,''
  Phys.\ Rev.\ D {\bf 41}, 1231 (1990).
  doi:10.1103/PhysRevD.41.1231


\bibitem{Nowakowski:2004cv} 
  M.~Nowakowski, E.~A.~Paschos and J.~M.~Rodriguez,
  ``All electromagnetic form-factors,''
  Eur.\ J.\ Phys.\  {\bf 26}, 545 (2005)
  doi:10.1088/0143-0807/26/4/001
  [physics/0402058].


\bibitem{Giunti:2008ve} 
  C.~Giunti and A.~Studenikin,
  ``Neutrino electromagnetic properties,''
  Phys.\ Atom.\ Nucl.\  {\bf 72}, 2089 (2009)
  doi:10.1134/S1063778809120126
  [arXiv:0812.3646 [hep-ph]].


\bibitem{Broggini:2012df} 
  C.~Broggini, C.~Giunti and A.~Studenikin,
  ``Electromagnetic Properties of Neutrinos,''
  Adv.\ High Energy Phys.\  {\bf 2012}, 459526 (2012)
  doi:10.1155/2012/459526
  [arXiv:1207.3980 [hep-ph]].


\bibitem{Giunti:2015gga} 
  C.~Giunti, K.~A.~Kouzakov, Y.~F.~Li, A.~V.~Lokhov, A.~I.~Studenikin and S.~Zhou,
  ``Electromagnetic neutrinos in laboratory experiments and astrophysics,''
  Annalen Phys.\  {\bf 528}, 198 (2016)
  doi:10.1002/andp.201500211
  [arXiv:1506.05387 [hep-ph]].


\bibitem{Velo:1969bt} 
  G.~Velo and D.~Zwanziger,
  ``Propagation and quantization of Rarita-Schwinger waves in an external electromagnetic potential,''
  Phys.\ Rev.\  {\bf 186}, 1337 (1969).
  doi:10.1103/PhysRev.186.1337


  \bibitem{Madore}
  J.~Madore,
  ``The characteristic surfaces of a classical spin-3/2 field in an Einstein-Maxwell background." 
  Phys.\ Lett.\ B {\bf 55.2}, 217 (1975).
  doi:10.1016/0370-2693(75)90446-3
  
  \bibitem{MadoreTait}
  J.~Madore and W.~Tait,
  ``Propagation of shock waves in interacting higher spin wave equations." 
  Comm.\ Math.\ Phys.\ {\bf 30.3} 201 (1973).
  doi:10.1007/BF01837358

\bibitem{Deser:2001dt} 
  S.~Deser and A.~Waldron,
  ``Inconsistencies of massive charged gravitating higher spins,''
  Nucl.\ Phys.\ B {\bf 631}, 369 (2002)
  doi:10.1016/S0550-3213(02)00199-2
  [hep-th/0112182].


\bibitem{Rahman:2011ik} 
  R.~Rahman,
  ``Helicity-1/2 mode as a probe of interactions of a massive Rarita-Schwinger field,''
  Phys.\ Rev.\ D {\bf 87}, no. 6, 065030 (2013)
  doi:10.1103/PhysRevD.87.065030
  [arXiv:1111.3366 [hep-th]].


\bibitem{Deser:2012qx} 
  S.~Deser and A.~Waldron,
  ``Acausality of Massive Gravity,''
  Phys.\ Rev.\ Lett.\  {\bf 110}, no. 11, 111101 (2013)
  doi:10.1103/PhysRevLett.110.111101
  [arXiv:1212.5835 [hep-th]].


\bibitem{Deser:2014hga} 
  S.~Deser, M.~Sandora, A.~Waldron and G.~Zahariade,
  ``Covariant constraints for generic massive gravity and analysis of its characteristics,''
  Phys.\ Rev.\ D {\bf 90}, no. 10, 104043 (2014)
  doi:10.1103/PhysRevD.90.104043
  [arXiv:1408.0561 [hep-th]].


\bibitem{D'Amico:2012zv} 
  G.~D'Amico, G.~Gabadadze, L.~Hui and D.~Pirtskhalava,
  ``Quasidilaton: Theory and cosmology,''
  Phys.\ Rev.\ D {\bf 87}, 064037 (2013)
  doi:10.1103/PhysRevD.87.064037
  [arXiv:1206.4253 [hep-th]].

\bibitem{deRham:2010kj} 
  C.~de Rham, G.~Gabadadze and A.~J.~Tolley,
  ``Resummation of Massive Gravity,''
  Phys.\ Rev.\ Lett.\  {\bf 106}, 231101 (2011)
  doi:10.1103/PhysRevLett.106.231101
  [arXiv:1011.1232 [hep-th]].

\bibitem{Adams:2006sv} 
  A.~Adams, N.~Arkani-Hamed, S.~Dubovsky, A.~Nicolis and R.~Rattazzi,
  ``Causality, analyticity and an IR obstruction to UV completion,''
  JHEP {\bf 0610}, 014 (2006)
  doi:10.1088/1126-6708/2006/10/014
  [hep-th/0602178].


\bibitem{Cheung:2016drk} 
  C.~Cheung, K.~Kampf, J.~Novotny, C.~H.~Shen and J.~Trnka,
  ``A Periodic Table of Effective Field Theories,''
  arXiv:1611.03137 [hep-th].
  

\end{thebibliography}
\end{document}